\documentclass[letterpaper]{article} 
\usepackage[]{aaai2026}  
\usepackage{times}  
\usepackage{helvet}  
\usepackage{courier}  
\usepackage[hyphens]{url}  
\usepackage{graphicx} 
\usepackage{natbib}  
\usepackage{caption} 
\usepackage{bm}
\usepackage{graphicx}
\usepackage{float}
\usepackage[utf8]{inputenc}
\usepackage{enumitem}
\usepackage{amsfonts}
\usepackage{algorithm, algpseudocode}
\usepackage{pgfplots}
\usepackage{makecell}
\usepackage{arydshln}
\usepackage{subcaption}
\usepackage{tikz}
\usepackage{adjustbox}
\usepackage{booktabs}
\usepackage{threeparttable}
\usepackage{comment}
\usepackage{mathtools}
\usepackage[marginal]{footmisc}

\usepackage{multirow}
\usepackage{amsmath}
\usepackage{amssymb}
\usepackage{tabularx}

\usepackage{enumitem}
\usepackage{booktabs}
\usepackage{newfloat}
\usepackage{listings}
\DeclareCaptionStyle{ruled}{labelfont=normalfont,labelsep=colon,strut=off} %
\floatstyle{ruled}
\newfloat{listing}{tb}{lst}{}
\floatname{listing}{Listing}

\title{Privacy on the Fly: A Predictive Adversarial Transformation Network for Mobile Sensor Data
}
\author {
    Tianle Song\textsuperscript{\rm 1},
    Chenhao Lin\textsuperscript{\rm 1}\thanks{Corresponding author.},
    Yang Cao\textsuperscript{\rm 2},
    Zhengyu Zhao\textsuperscript{\rm 1},
    Jiahao Sun\textsuperscript{\rm 1},
    Chong Zhang\textsuperscript{\rm 1},
    Le Yang\textsuperscript{\rm 1},
    and Chao Shen\textsuperscript{\rm 1}
}

\affiliations {
    \textsuperscript{\rm 1}Xi’an Jiaotong University, China  \\
    \textsuperscript{\rm 2}Institute of Science Tokyo, Japan \\
    \{tianlesong, sunjiahao\}@xjtu.stu.edu.cn, 
    \{linchenhao, zhengyu.zhao, zhangchong, yangle15, chaoshen\}@xjtu.edu.cn, \\
    cao@c.titech.ac.jp
    }

\usepackage{fancyhdr}
\usepackage{footmisc}   

\usepackage{bibentry}
\makeatletter
\def\copyright@on{F}
\makeatother
\begin{document}
\maketitle

\begin{abstract}

Mobile motion sensors such as accelerometers and gyroscopes are now ubiquitously accessible by third-party apps via standard APIs. While enabling rich functionalities like activity recognition and step counting, this openness has also enabled unregulated inference of sensitive user traits, such as gender, age, and even identity, without user consent. Existing privacy-preserving techniques, such as GAN-based obfuscation or differential privacy, typically require access to the full input sequence, introducing latency that is incompatible with real-time scenarios. Worse, they tend to distort temporal and semantic patterns, degrading the utility of the data for benign tasks like activity recognition. To
address these limitations, we propose the Predictive Adversarial Transformation Network (PATN), a real-time privacy-preserving framework that leverages historical signals to generate adversarial perturbations proactively. The perturbations are applied immediately upon data acquisition, enabling continuous protection without disrupting application functionality. Experiments on two datasets demonstrate that PATN substantially degrades the performance of privacy inference models, achieving Attack Success Rate (ASR) of 40.11\% and 44.65\% (reducing inference accuracy to near-random) and increasing the Equal Error Rate (EER) from 8.30\% and 7.56\% to 41.65\% and 46.22\%. On ASR, PATN outperforms baseline methods by 16.16\% and 31.96\%, respectively.

\end{abstract}


\section{Introduction}

Mobile applications can easily access motion sensor data (e.g., accelerometer and gyroscope) through APIs on Android and iOS platforms~\cite{rajguru2019sensor}. This supports various functions like activity recognition~\cite{wang2019deep}, step counting~\cite{susi2013motion}, and gesture interaction~\cite{brotchie2022leveraging}, making motion data central to mobile services. However, this widespread access also raises privacy concerns, as sensor data can reveal personal attributes like identity~\cite{cai2024famos,fereidooni2023authentisense,shen2017performance}, gender~\cite{meena2020gender}, and age~\cite{lin2023childshield, miao2023learning}. Moreover, the availability of open-source privacy inference models~\cite{malekzadeh2019motionsense} makes it easier for third-party applications to exploit such data, potentially infringing on user privacy without their knowledge. These concerns underscore the need for mechanisms that can safeguard user privacy without compromising the utility of motion sensor data.

\begin{figure}[t!]
    \centering
    \includegraphics[width=0.47\textwidth]{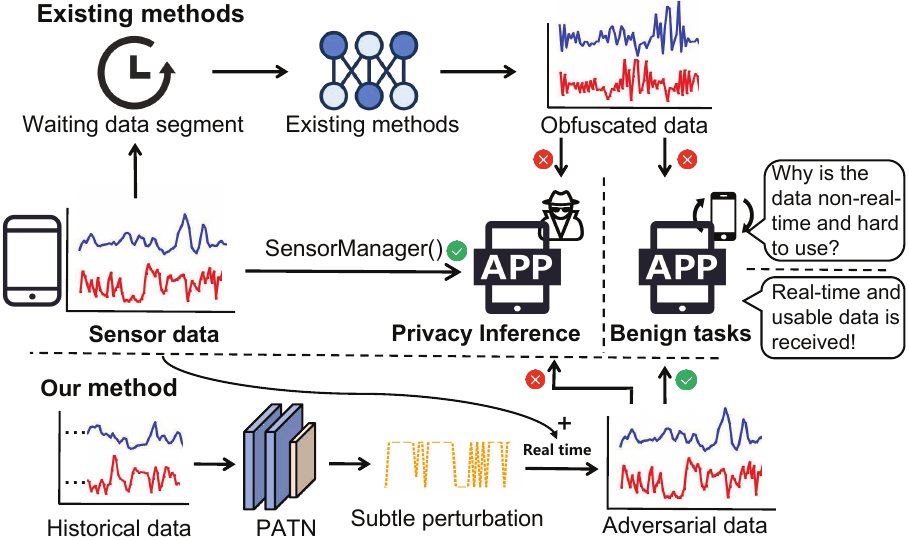}
    \caption{Our method vs. existing methods, addressing two issues: (1) temporal semantic distortion and (2) delayed obfuscation due to segment-wise processing. }
    \label{fig:problem1}
\end{figure}

Existing privacy-preserving approaches, including differential privacy~\cite{kalupahana2023serandip} and generative model–based obfuscation~\cite{raval2019olympus,gu2025towards}, often optimize for specific tasks to preserve utility while preventing private information inference. However, as shown in Figure~\ref{fig:problem1}, these methods face two key limitations: \textbf{(1)} Generative methods often compromise temporal semantics by regenerating entire sequences via latent-space sampling, over-smoothing or distorting fine-grained patterns, and degrading utility for tasks requiring precise numerical computations (e.g., smartphone rotation angle estimation). \textbf{(2)} Most cannot satisfy real-time demands, as they buffer complete sensor sequences before transformation, whereas real-world streams must be processed instantly. These challenges highlight the need for methods that protect sensitive information while maintaining both semantic fidelity and real-time usability.

Adversarial perturbations exploit model sensitivity by adding imperceptible noise that induces incorrect outputs. We adopt them as a lightweight way to suppress sensitive patterns while preserving sensor data semantics. Although adversarial attacks on time-series models~\cite{yang2022tsadv,pialla2025time} are well studied, most rely on full input segments, making them unsuitable for streaming and unable to address limitation (2). Simply applying perturbations generated from historical data directly to future inputs has limited effectiveness, as it may fail to align with the temporal dynamics of the target sequence. Universal adversarial perturbations~\cite{rathore2020untargeted} often perform poorly on complex, high-dimensional sensor data, further limiting their practicality.

To address these issues, we propose the \textit{Predictive Adversarial Transformation Network} (PATN), which leverages historical sensor data to generate adversarial perturbations for future readings. This enables real-time, zero-latency protection with high consistency to original signals. PATN combines a generative network with a \textit{history-aware top-$k$ optimization} strategy to mitigate temporal misalignment in inference attacks. Experiments on two datasets show PATN achieves 40.11\% and 44.65\% Attack Success Rate (ASR), reduces Equal Error Rate (EER) to 41.65\% and 46.22\%, and preserves semantic fidelity for non-sensitive applications.
The implementation of PATN is publicly available at \url{https://github.com/skysky4/PATN}.
Our contributions are summarized as follows,

\begin{itemize} 
\item  We propose PATN, the first framework to enable proactive, real-time privacy protection for streaming sensor data by generating future-directed adversarial perturbations, and the first to introduce adversarial perturbations into sensor data privacy protection.

\item  We introduce a history-aware top-$k$ optimization strategy that effectively counters performance degradation under temporal misalignment attacks.
\item  Through extensive evaluation, PATN is shown to substantially enhance privacy protection while retaining the usability and semantic fidelity of motion sensor signals.
\end{itemize}

\section{Related Works}

\noindent\textbf{Data Obfuscation.} Privacy-preserving methods for sensor data fall into three categories. Differential privacy (DP)~\cite{kalupahana2023serandip} adds noise to protect privacy but distorts temporal signals, reducing utility in continuous tasks. Privacy-aware feature extraction~\cite{liu2019privacy, li2021deepobfuscator} learns representations that retain task-relevant information while masking sensitive attributes, often via adversarial training. Recent generative approaches, such as GANs~\cite{boutet2021dysan}, VAEs~\cite{hajihassnai2021obscurenet}, and diffusion models~\cite{yang2023privacy}, produce realistic sensor data but often face semantic drift, affecting fine-grained temporal patterns. For example, DoppelGANger~\cite{lin2020using} captures long-term trends but struggles with accelerometer dynamics, while PrivDiffuser~\cite{yang2025privdiffuser} improves privacy-utility trade-offs but may distort temporally sensitive signals. Though generative methods obfuscate private information well, they risk degrading signal semantics, limiting use in sensor-driven applications. Additionally, existing methods cannot meet real-time demands, as they buffer entire sequences before transformation.

\noindent\textbf{Adversarial Attacks.} Adversarial attacks on time series have been extensively studied. Early work examined gradient-based attacks, including FGSM~\cite{fawaz2019adversarial,oregi2018adversarial} and PGD on deep time-series classifiers. Comprehensive surveys such as Adversarial Attacks on Time Series provide taxonomies of these methods and their impacts across various domains~\cite{karim2020adversarial}. Özcan et al. introduced AdaptEdge~\cite{khan2024adaptedge}, a targeted universal attack framework that demonstrated cross-dataset transferability in smart grid time series. Recent methods, such as SFAttack~\cite{gu2025towards} and BlackTreeS~\cite{ding2023black}, targeting imperceptible adversarial attacks for time series classification, improve stealth by utilizing localized and frequency-domain perturbations. While these studies highlight the vulnerability of time-series models, they uniformly assume access to complete time-series segments before perturbation, which limits their applicability to streaming sensor data that must be protected in real-time.

\begin{figure}[t!]
    \centering
    \includegraphics[width=0.45\textwidth]{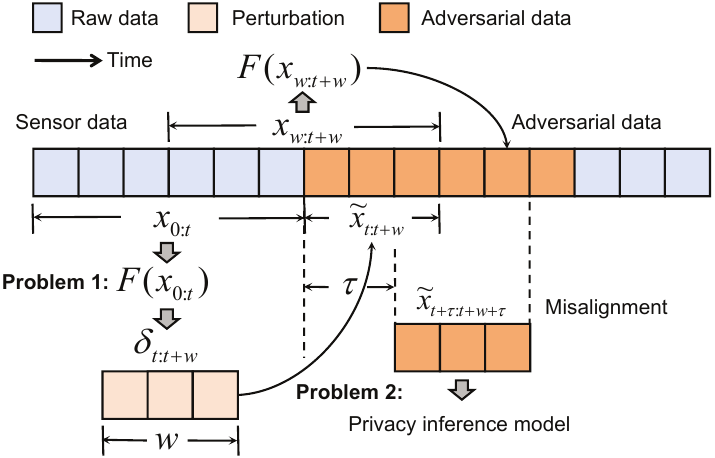}
    \caption{Illustrating the two problems: generating real-time perturbations for continuously arriving sensor data, and addressing temporal misalignment where attacks may occur at arbitrary, unpredictable time points.}
    \label{fig:problem}
\end{figure}

\begin{figure*}[t!]
    \centering
    \includegraphics[width=0.85\textwidth]{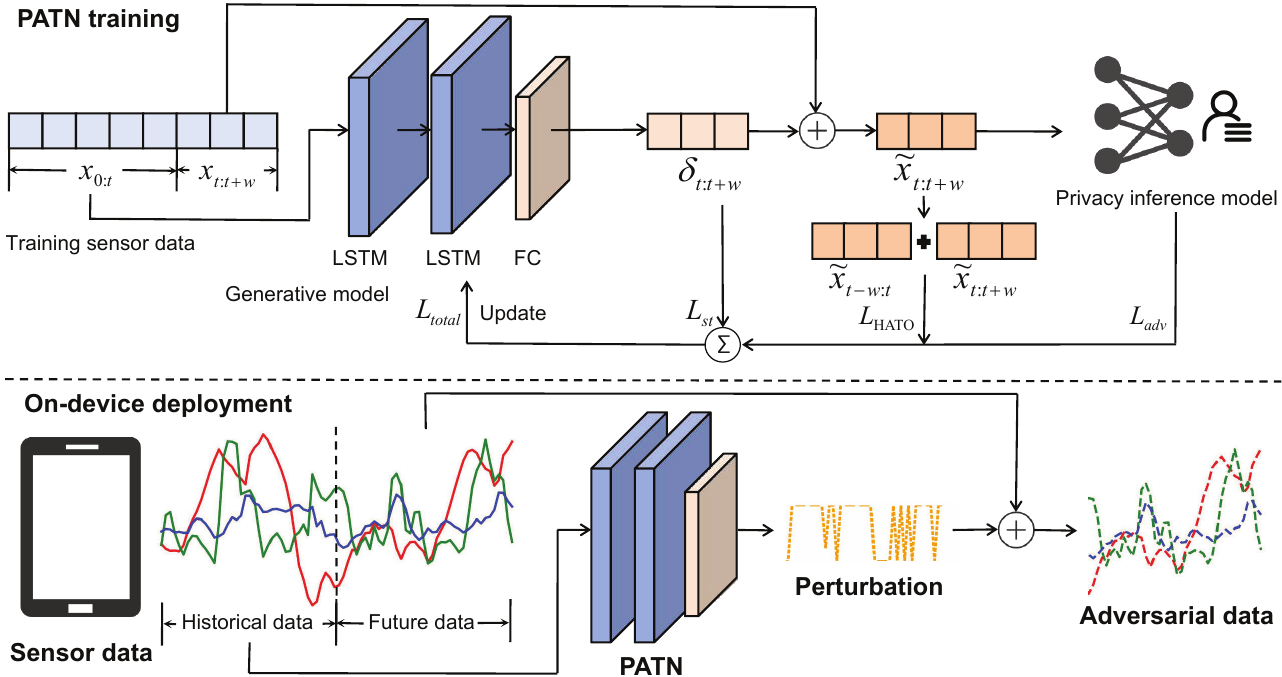}
    \caption{Overview of the PATN framework. The system includes PATN training, where the network learns to generate privacy-preserving perturbations, and on-device deployment, where the trained model runs securely in real time.}
    \label{fig:oveview}
\end{figure*}

\section{Problem Setting}

Our objective is to protect user privacy by adding adversarial perturbations to raw sensor data while preserving the semantic integrity of the original signals. Based on this objective, we identify two key problems.

\noindent\textbf{Problem 1: Real-Time Perturbation Generation.}
The problem we need to solve is to generate real-time, future-directed adversarial perturbations to protect user privacy from inference attacks on motion sensor data. Existing methods typically require access to complete sensor sequences before adding perturbations or reconstructing data, making them incompatible with real-time applications where sensor data arrive continuously and must be processed on the fly.

As shown in Figure~\ref{fig:problem}, we learn a temporal mapping function \( \mathcal{F} \) that takes historical motion data \( x_{0:t} \) as input and outputs a sequence of perturbations for the upcoming steps:
\begin{equation}
\delta_{t:t+w} = \mathcal{F}(x_{0:t})
\end{equation}
The predicted perturbations \( \delta_{t:t+w} \) are then applied in real time to the incoming data \( x_{t}, x_{t+1}, \ldots \), ensuring that privacy-sensitive patterns are obfuscated before being accessed by untrusted applications.

\noindent\textbf{Problem 2: Temporal Misalignment Between Defense and Attack.}
The second problem is ensuring defense effectiveness under temporal misalignment (Figure~\ref{fig:problem}). In practice, adversaries may launch inference attacks at arbitrary time points, while existing defenses generate perturbations aligned to fixed timelines, reducing their effectiveness. Let \( \delta_{t:t+w} \) denote the perturbation over window \([t, t+w]\), and suppose the attacker targets a shifted window \( \tilde{x}_{t+\tau:t+\tau+w} \) with unknown offset \( \tau \), such that \( \tilde{x}_{t:t+w} \neq \tilde{x}_{t+\tau:t+\tau+w} \). This misalignment causes partial overlap between the perturbation and attack windows, weakening privacy protection.

\section{Methodology}

\subsection{Overview}

To protect user privacy while preserving sensor data semantics, we propose the PATN framework, as shown in Figure~\ref{fig:oveview}. Assuming access to open-source privacy inference models and their gradients (white-box), PATN leverages historical sensor data to predict future-directed perturbations, optimized to mislead privacy inference models with minimal semantic distortion. The framework has two components: (1) \textit{PATN training}, where PATN is optimized using three objectives—adversarial effectiveness, temporal robustness, and smoothness regularization—to balance privacy protection and data fidelity; and (2) \textit{On-device deployment}, where the trained network is executed locally on mobile devices to enable zero-latency perturbation generation for real-time sensor streams.

\subsection{Predictive Adversarial Transformation Network}
\subsubsection{Perturbation Range Setting.}

To prevent adversarial perturbations from degrading the semantic integrity of sensor data or disrupting normal system functionality, we constrain their magnitude with carefully designed bounds based on $\ell_{\infty}$ normalization. Excessively small perturbations risk ineffective attacks, while overly large ones may distort essential patterns.

We first derive a dimension-wise upper bound based on dataset statistics by computing the mean and standard deviation for each sensor dimension and defining the perturbation limit as 0.05 of these values:
\begin{equation}
\epsilon_d^{\text{data}} = \min(0.05 \times \mu_d,\ 0.05 \times \sigma_d)
\end{equation}
where $\mu_d$ and $\sigma_d$ denote the mean and standard deviation of the $d$-th dimension.

To further ensure imperceptibility, we measure natural sensor variation under static conditions, where 10 users interacted with smartphones fixed on a rigid table. The standard deviation of each feature dimension in this scenario provides $\epsilon_d^{\text{natural}}$, which we combine with the statistical bound to define the final $\ell_{\infty}$ constraint:
\begin{equation}
\epsilon_d = \min(\epsilon_d^{\text{data}},\ \epsilon_d^{\text{natural}})
\end{equation}
This guarantees that perturbations remain within naturally occurring fluctuations, preserving utility while ensuring privacy.

\subsubsection{History-Driven Generative Model for Perturbation.}

To address the challenge of real-time perturbation generation, we propose PATN, a generative framework that produces future adversarial perturbations based solely on historical multivariate sensor data. Unlike conventional adversarial methods that require future observations, PATN forecasts perturbations from past system sensor data, making it suitable for real-world time-series scenarios,  where future data is unavailable at inference time.

PATN utilizes a sequence-to-sequence architecture with an LSTM-based encoder-decoder framework. The encoder processes a multivariate input sequence of length $T_{\text{in}}$, consisting of $D$-dimensional sensor readings, denoted as $x \in \mathbb{R}^{B \times T_{\text{in}} \times D}$. The LSTM encoder extracts temporal dependencies and condenses the sequence into a fixed-length latent representation, typically the final hidden and cell states. The decoder, another LSTM, autoregressively generates a sequence of adversarial perturbations of length $T_{\text{out}}$, with each step producing a perturbation vector $\delta_i \in \mathbb{R}^D$.
\begin{equation}
    \delta_i = W_o h_i + b_o
\end{equation}
Here, $W_o \in \mathbb{R}^{D \times H}$ and $b_o \in \mathbb{R}^D$ are learnable parameters, and $H$ denotes the hidden dimension of the decoder LSTM. The decoder operates in an autoregressive manner, meaning the output from each previous step can be fed back as input into the next, allowing the network to capture temporal consistency in the generated perturbation sequence.
The complete adversarial sequence is defined as:
    \begin{equation}
        \delta = [\delta_1, \delta_2, \dots, \delta_{T_{\text{out}}}] \in \mathbb{R}^{B \times T_{\text{out}} \times D}, \|\delta\|_\infty \leq \epsilon_d
    \end{equation}
This predicted perturbation tensor is subsequently added to the raw future sensor data to construct adversarial examples:
\begin{equation}
    \tilde{x}_{\text{adv}} = x_{\text{raw}} + \delta
\end{equation}
where $x_{\text{raw}} \in \mathbb{R}^{B \times T_{\text{out}} \times D}$ denotes the future clean data sequence and $\tilde{x}_{\text{adv}}$ is the resulting adversarial input.

\subsubsection{History-Aware Top-k Optimization.}
\label{History}

\begin{algorithm}[t!]
\caption{History-Aware Top-k  Optimization}
\label{alg:history_adv_topk}
\begin{algorithmic}[1]
\State \textbf{Input:} Previous perturbation $\delta_{t-w:t}$, current perturbation $\delta_{t:t+w}$, input $x $, model $f$, label  \( y_{\text{target}} \) , window size $w$, step size $s$, top-k $k$
\State \textbf{Output:} Top-k averaged loss 

\State Merge perturbations: $\delta \gets \text{Merge}(\delta_{t-w:t}, \delta_{t:t+w})$
\State Generate adversarial input: $\tilde{x} \gets x + \delta$

\State Initialize loss list: $\mathcal{L} \gets [\,]$

\For{each $t \in \{0, s, 2s, \dots, T - w\}$}
    \State Extract window: $\tilde{x}_t \gets \tilde{x}[:, t : t+w, :]$
    \State Compute prediction: $\hat{z}_t \gets f(\tilde{x}_t)$
    \State Compute loss: $\ell_t \gets \text{CrossEntropy}(\hat{z}_t,\  y_{\text{target}} )$
    \State Append loss: $\mathcal{L}.\text{append}(\ell_t)$
\EndFor

\State Stack all losses: $\ell_{\text{stack}} \gets \text{stack}(\mathcal{L})$
\State Select top-$k$ losses: $\ell_{\text{topk}} \gets \text{TopK}(\ell_{\text{stack}},\ k)$
\State Compute final loss: $\mathcal{L}_{\text{HATO}} \gets \text{mean}(\ell_{\text{topk}})$

\State \Return $\mathcal{L}_{\text{HATO}}$
\end{algorithmic}
\end{algorithm}

To enhance the robustness of adversarial perturbations against Problem 2, we propose history-aware top-k optimization (HATO). This method leverages both the previously generated perturbation (the last perturbation segment) and the current perturbation, aiming to construct a more temporally consistent and generalizable perturbation sequence.

As shown in Algorithm~\ref{alg:history_adv_topk}, we first concatenate the previous perturbation $\delta_{\text{prev}}$ with the current perturbation $\delta_{\text{cur}}$ to form a longer adversarial perturbation. This merged perturbation is then applied to the clean input sequence. A sliding window with a fixed size and stride is used to extract overlapping temporal segments. For each segment, the privacy inference model $f$ computes a cross-entropy loss with a misleading target label to simulate a privacy-preserving adversarial objective.
To address the instability of loss values across different window positions, we adopt a top-$k$ selection strategy. Among the computed losses, we select the $k$ highest values and use their average as the final optimization target.
\begin{equation}
\mathcal{L}_{\text{HATO}} = \frac{1}{k} \sum_{i=1}^{k} \text{TopK}_i(\mathcal{L}, k)
\label{hitop}
\end{equation}
This encourages the perturbation to consistently degrade model performance across multiple subwindows, rather than overfitting to a specific temporal slice.
This history-aware top-k loss not only improves the temporal robustness of the perturbation but also enhances its defensive effect under arbitrary window-based evaluation, which is common in real-world privacy-preserving scenarios.

\subsubsection{Objective Function.}

To optimize the PATN model, we define a composite loss function that integrates three complementary objectives.

1. \textbf{Misclassification loss} \( \mathcal{L}_{\text{adv}} \): This term encourages the adversarial input to mislead the target model (Privacy inference model). It is implemented as a standard cross-entropy loss between the model's prediction and a misleading target:
\begin{equation}
\mathcal{L}_{\text{adv}} = \mathcal{L}_{\text{CE}}(\text{logits}_{\text{adv}},\ y_{\text{target}})
\end{equation}
where \( y_{\text{target}} \) denotes a misleading label, which in a binary classification task corresponds to the opposite of the label output by the privacy inference model.

2. \textbf{History-aware top-$k$ loss} \( \mathcal{L}_{\text{HATO}} \): As detailed in Equation~\ref{hitop}, this loss promotes temporal robustness by encouraging consistent model outputs across a sliding inference window. This helps the perturbation remain effective even under windowed decision-making.

3. \textbf{Smoothness loss} \( \mathcal{L}_{\text{st}} \): This regularization term penalizes abrupt changes in the perturbation signal over time, thereby improving the visual and temporal coherence of the generated adversarial sequence. It is implemented as the mean squared error (MSE) of the perturbation:
\begin{equation}
\mathcal{L}_{\text{st}} = \text{MSE}(\delta)
\end{equation}

The total objective is a weighted sum:
\begin{equation}
\mathcal{L}_{\text{total}} = \mathcal{L}_{\text{adv}} + \lambda_1 \mathcal{L}_{\text{HATO}} + \lambda_2 \mathcal{L}_{\text{st}}
\end{equation}
with \( \lambda_1 = 0.3 \), \( \lambda_2 = 0.3 \). These weights are chosen to balance attack effectiveness, temporal robustness, and perceptual imperceptibility. When confronted with multiple privacy inference models, we aggregate the $\mathcal{L}_{\text{total}}$ from each model into a unified optimization objective, enabling the method to effectively defend against a diverse set of attacks simultaneously.

\begin{table*}[t!]
\centering  
\setlength{\tabcolsep}{8.8pt}  
\begin{tabular}{c|cccc|cccc}
\hline
\multirow{2}{*}{Method} & \multicolumn{4}{c|}{MotionSense}                              & \multicolumn{4}{c}{ChildShield}                               \\ 
                        & ASR(\%)$\uparrow$                           & EER(\%)$\uparrow$  & AUC$\downarrow$    & F1\_Score$\downarrow$    & ASR(\%)$\uparrow$                           & EER(\%)$\uparrow$  & AUC$\downarrow$     & F1\_Score$\downarrow$    \\ \hline
Raw data                & \textbackslash{} & 8.30    & 0.979 & 0.928    & \textbackslash{} & 7.56    & 0.972 & 0.922   \\
DP                      & 14.37                            & 17.46   & 0.931 & 0.836    & 4.12                            & 12.00   & 0.952 & 0.887    \\
UAP                     & 9.61                             & 13.53    & 0.951 & 0.863    & 3.17                            & 10.92    & 0.957 & 0.894    \\
FGSM                    &23.95                              & 25.92    & 0.772 & 0.802    & 12.99                          & 19.11      & 0.889 & 0.821      \\
PGD                     &23.95                              & 25.92    & 0.772 & 0.802   &  12.99                          & 19.11      & 0.889 & 0.821   \\
\textbf{PATN(ours)}               & \textbf{40.11}                            & \textbf{41.65}   & \textbf{0.662} & \textbf{0.611}    & \textbf{44.95}                            & \textbf{46.22}   & \textbf{0.549} & \textbf{0.537}    \\ \hline
\end{tabular}
\caption{Performance comparison with baseline methods.}
\label{tab:performance_comparison}
\end{table*}  
  
\section{Experiments}

In this section, we describe the experimental setup and evaluate our method via baseline comparisons, utility analysis, ablation studies, and black-box migration tests to assess performance, component contributions, and robustness.

\subsection{Implementation Details}

\subsubsection{Datasets.}



We evaluate our method on two real-world datasets: MotionSense~\cite{Malekzadeh:2018:PSD:3195258.3195260} and ChildShield~\cite{lin2023childshield}. Both provide accelerometer and gyroscope data, split 7:3 for training and testing.

MotionSense contains 50Hz motion data from 24 participants using an iPhone 6s during six activities (walking, jogging, sitting, standing, upstairs, and downstairs), enabling gender (male/female) inference.
ChildShield includes 60Hz motion data from 1,875 participants across five mobile games, used for age (child/adult) inference.

\subsubsection{Evaluation Metric.}


To evaluate privacy protection, we define the attack success rate (ASR) as the proportion of originally correct predictions by the privacy inference model that are misclassified after applying our perturbation. A higher ASR indicates stronger privacy protection. We also report Equal Error Rate (EER), Area Under the Curve (AUC), and F1\_Score to assess the overall impact on adversarial inference performance. For utility evaluation, we adopt two perspectives. First, we measure semantic consistency using $\ell_{2}$ distance, Dynamic Time Warping (DTW), low-frequency (LF) component, and Root Mean Square Error (RMSE)~\cite{wu2022small}, where lower values indicate better preservation of signal characteristics. Second, we assess the impact on downstream tasks, specifically step detection~\cite{susi2013motion} and human activity recognition (HAR)~\cite {mekruksavanich2021lstm}, to quantify the effect of perturbations on typical usage scenarios.

\subsubsection{Privacy Inference Models and Baseline Methods.}
Most privacy inference models for sensor data adopt CNN architectures. In this work, we examine two representative CNN-based models for extracting sensitive user information, following the architectures of Lin et al.~\cite{lin2023childshield} (age) and Sharshar et al.~\cite{sharshar2021activity} (gender). We also evaluate PATN on sequential models (RNN, LSTM), with results reported in the Appendix~\cite{PATN2025}.

We compare our approach with representative baselines under the same $\ell_{\infty}$-norm constraint. Data-independent methods include differential privacy (DP)~\cite{kalupahana2023serandip} and universal adversarial perturbations (UAP)~\cite{rathore2020untargeted}, while data-dependent methods include FGSM~\cite{fawaz2019adversarial} and PGD~\cite{oregi2018adversarial}, both leveraging historical sensor data. Generative model–based approaches~\cite{yang2025privdiffuser} are excluded from direct comparison, as they severely violate the $\ell_{\infty}$ constraint and require full sensor sequences at the current time step, which limits real-time applicability. Nonetheless, we qualitatively compare with PrivDiffuser~\cite{yang2025privdiffuser}, showing that our method better preserves semantic consistency in perturbed sensor signals (in section~\ref{ul}: Utility Analysis).


\subsubsection{Parameters.}  
The Predictive Adversarial Transformation Network (PATN) takes an input sequence of length $T_{\text{in}}$ = 30 (each representing 0.5-second intervals of sensor data) and generates an output sequence of length $T_{\text{out}}$ = 10, corresponding to the input length for the privacy inference model. The input dimension is 6, representing six sensor features, and the LSTM hidden dimension is 64. HATO is applied with a top-k parameter of k=2.
Training and testing are performed on an online server with an NVIDIA 3090Ti GPU using PyTorch. The Adam optimizer is used, with an initial learning rate of 1e-3 and a fixed-step decay schedule (halving every 200 epochs), over a total of 600 epochs.

\subsection{Main Results}

\begin{figure}[t!]
    \centering
    \includegraphics[width=0.46\textwidth]{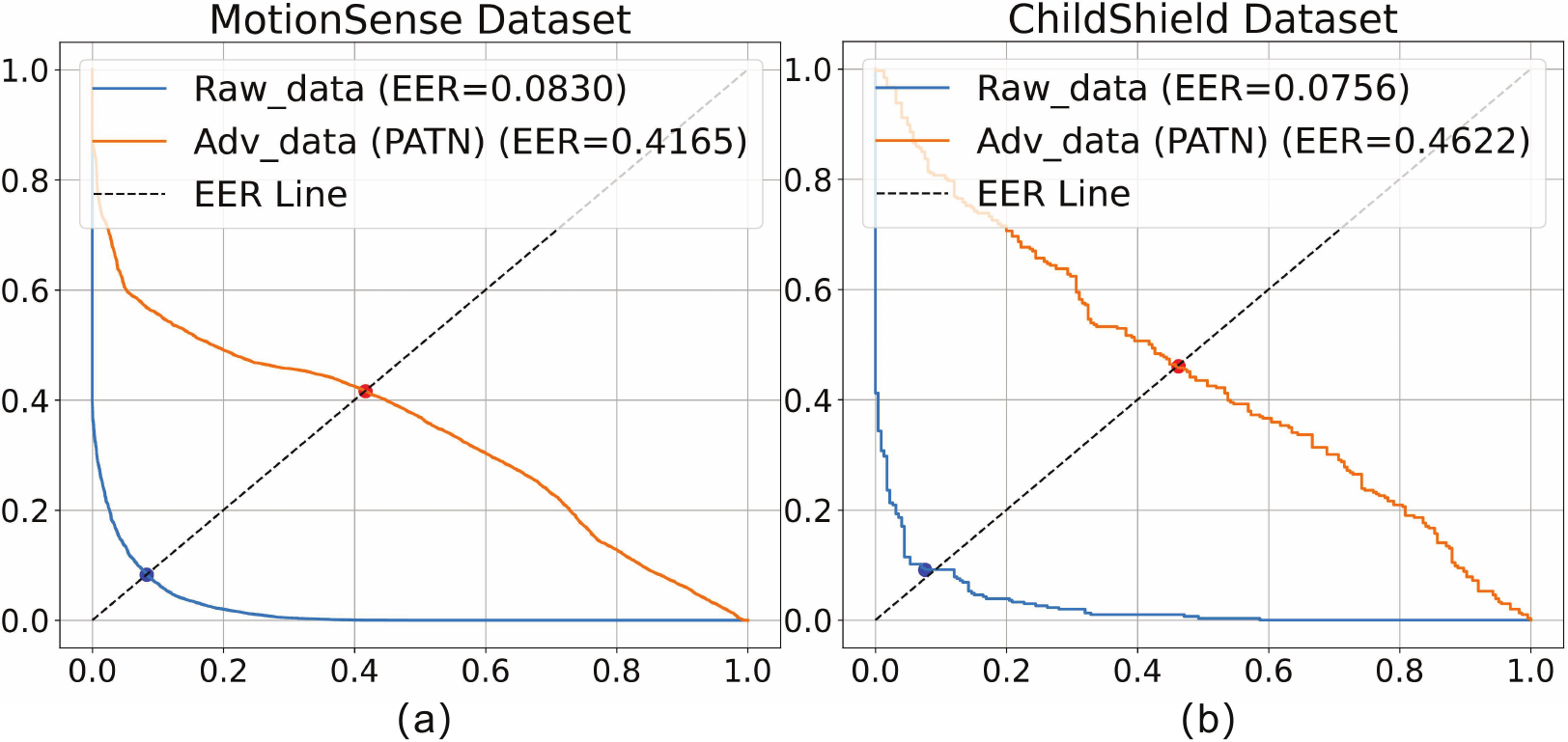}  
    \caption{The AUC curve of adversarial data generated by PATN compared to the raw data when applied to the privacy inference model.}
    \label{fig:mainres}
\end{figure}

\subsubsection{Comparison with Baseline Methods.}
As shown in Table \ref{tab:performance_comparison} and Figure~\ref{fig:mainres}, under the same $\ell_{\infty}$-norm constraint, our proposed PATN achieves superior privacy protection by generating adversarial perturbations that effectively mislead privacy inference models, targeting gender inference on MotionSense and age inference on ChildShield.

In contrast to traditional approaches like DP and UAP, PATN is dynamic and adaptive, optimizing perturbations specifically to protect privacy. DP and UAP rely on fixed, data-agnostic perturbation strategies that overlook the vulnerabilities of historical data distributions and fail to target real-time or future private attribute inference. PATN, by introducing a predictive adversarial transformation, tailors perturbations to the model’s current predictions and adapts them for future data, making them more effective in confusing privacy inference models. Consequently, PATN achieves substantially higher ASR 40.11\% on MotionSense and 44.95\% on ChildShield compared to DP (14.37\% and 4.12\%) and UAP (9.61\% and 3.17\%).

\begin{figure}[t!]
    \centering
    \includegraphics[width=0.47\textwidth]{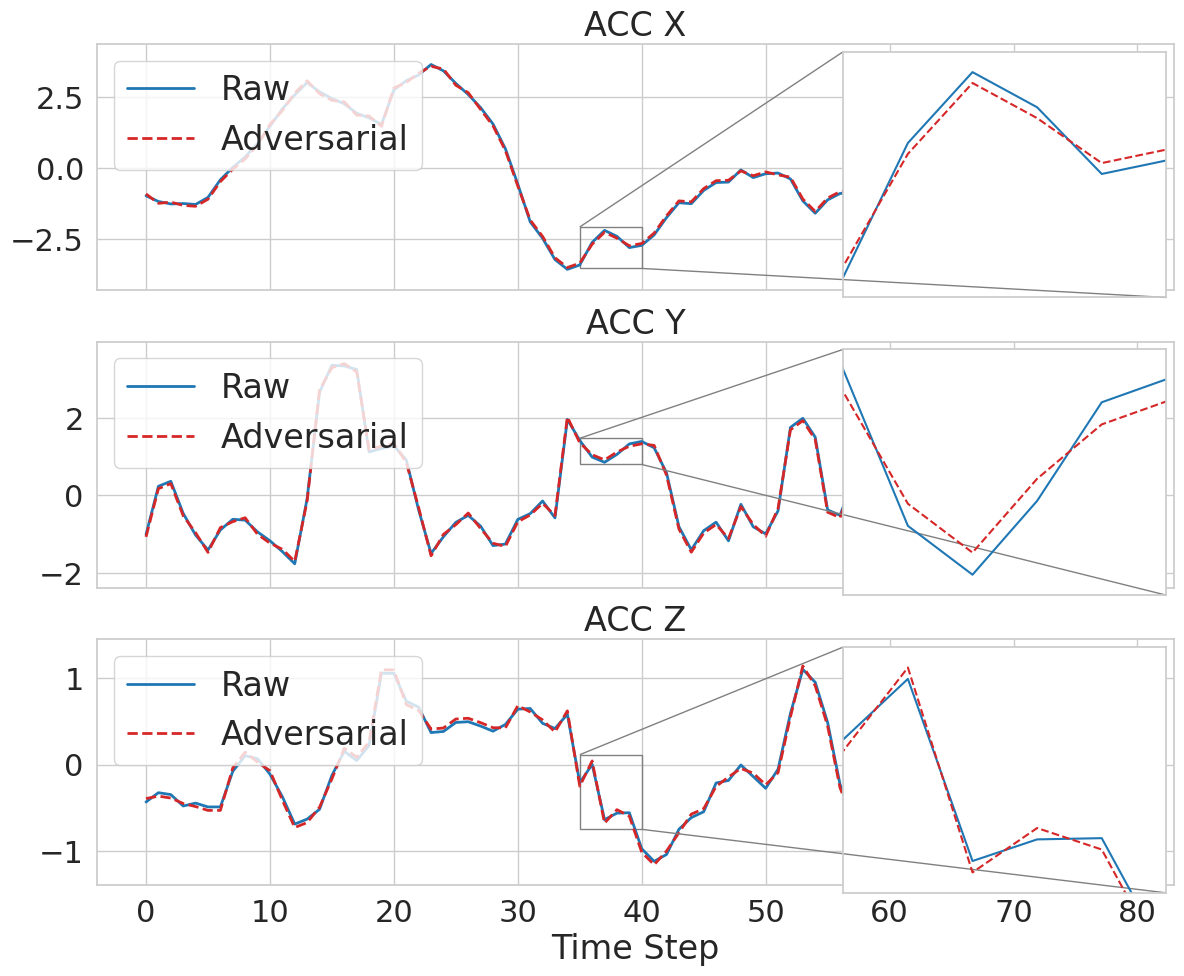}  
    \caption{Comparison of Raw and Adversarial sensor data (accelerometer) on IMU Axes.}
    \label{fig:sensoracc}
\end{figure}


While FGSM and PGD craft perturbations based on historical data distributions, they lack optimization for future data (i.e., unseen data), limiting their effectiveness in dynamic environments. These methods assume minimal temporal variation in sensor data, which constrains their robustness. As shown in Table \ref{tab:performance_comparison}, FGSM and PGD achieve ASRs of 23.95\% on MotionSense and 12.99\% on ChildShield, indicating partial success in misleading privacy inference models but without future adaptability.
In contrast, PATN explicitly optimizes perturbations for future data, enhancing robustness and privacy protection. It achieves significantly higher ASRs of 40.11\% and 44.95\% on MotionSense and ChildShield, respectively, demonstrating superior performance in scenarios with evolving data distributions.

\subsubsection{Utility Analysis.} 
\label{ul}

Our PATN model has a compact size of only 0.365 MB, well within the 2 MB memory constraint for deployment in the TEE. The perturbation generation time is just 0.00036 seconds, much faster than the 1/60-second sensor sampling interval, ensuring real-time applicability.

To assess semantic fidelity, we visualize the raw and perturbed accelerometer signals along individual IMU axes (Figure~\ref{fig:sensoracc}, gyroscope shown in Appendix~\cite{PATN2025}).
 The adversarial signals show only subtle deviations from the original, indicating minimal impact on the sensor’s temporal semantics, which is crucial for downstream tasks like step detection.
We also compute the per-segment difference between raw and adversarial signals in 10-second segments. Table~\ref{tab1} compares PATN with PrivDiffuser across four metrics. PATN achieves a DTW of 0.744, $\ell_{2}$ distance of 0.162, LF difference of 0.300, and RMSE of 0.037, all significantly lower than PrivDiffuser, confirming that PATN better preserves signal fidelity while obfuscating sensitive information.

\begin{table}[]
\centering
\setlength{\tabcolsep}{8.2pt} 
\begin{tabular}{c|cccc}
\hline
     & DTW$\downarrow$                   & $\ell_{2}$$\downarrow$  & LF$\downarrow$  & RMSE$\downarrow$ \\ \hline
PATN         & 0.744 &  0.162  &  0.300  & 0.037\\
PrivDiffuser &   7.058      &   2.251 &   3.422 & 0.503\\ \hline
\end{tabular}
\caption{Comparison of Semantic Consistency Between PATN and PrivDiffuser}
\label{tab1}
\end{table}

\begin{table}[!t]
    \centering
    \setlength{\tabcolsep}{3.1pt} 
    \begin{tabular}{c|ccc}
    \hline
        ~ & Raw data & PrivDiffuser & PATN \\ \hline
        Step detection & 7916& 8683 (+767) & 7937 (+21)    \\ 
        HAR-EER(\%)\textsuperscript{1} & 5.08 &6.92 (+1.84) & 6.57 (+1.49)\\ 
HAR-AUC\textsuperscript{2} & 0.987 & 0.976 (-0.011) & 0.981 (-0.006)\\
\hline
    \end{tabular}
\begin{tablenotes}
\scriptsize
\item \textsuperscript{1}    HAR-EER(\%): human activity recognition (Macro-EER)
\item \textsuperscript{2}    HAR-AUC: human activity recognition (Macro-AUC)
\end{tablenotes}
\caption{Comparison of Raw Data, PrivDiffuser and PATN-perturbed Data in Step Detection and HAR}
\label{tab:HAR}
\end{table}



We conduct experiments on the MotionSense dataset, focusing on walking data for step detection and multi-class human activity recognition (HAR). As shown in Table~\ref{tab:HAR}, our method applies subtle $\ell_{\infty}$-bounded perturbations that preserve sensor utility. In step detection, PATN leads to only a negligible increase of 21 steps, showing minimal impact on usage. For HAR, it introduces a small change in EER (+1.49\%) and AUC (–0.006), indicating real-time activity inference remains viable. 

In contrast, PrivDiffuser, although optimized for HAR task (EER: 6.92\%, AUC: 0.976), shows poor generalization to other benign tasks, causing a large step count error (+767 steps). This highlights PATN’s task-agnostic design, achieving a better balance between privacy and fidelity without task-specific tuning.



\begin{table}[t!]
\centering
\setlength{\tabcolsep}{9pt} 
\begin{tabular}{c|ccc}
\hline
Model      & CNN  & ResNet & DenseNet \\ \hline
ASR(\%)        & 36.98 &  38.09  &  38.64 \\
Raw data EER(\%) &   8.30      &   7.36 &   7.85 \\
Adv data EER(\%) &   39.64      &   38.66 &   40.20\\ \hline
\end{tabular}
\caption{Evaluation of PATN’s effectiveness across multiple privacy inference models. }
\label{tab:patn_results}
\end{table}

\subsubsection{Multiple Privacy Inference Models.}
\label{mumodel}

In real-world scenarios, multiple privacy inference models may simultaneously threaten user privacy, so a method like PATN must remain robust against diverse attack strategies. We evaluate PATN on the MotionSense dataset (used for all subsequent experiments) against three CNN architectures: a standard CNN, a ResNet, and a DenseNet. PATN is jointly optimized against all three models during training to ensure comprehensive defense.

The results, summarized in Table~\ref{tab:patn_results}, show that PATN effectively mitigates attacks across these architectures. Specifically, PATN achieves ASR from 36.98\% to 38.64\% and EER between 38.64\% and 40.20\%, demonstrating consistent defense capability across different privacy inference models. These findings highlight PATN's versatility and robustness in protecting sensor data privacy.

\subsection{Ablation Studies}   

\subsubsection{The Impact of Different Input Lengths $T_{\text{in}}$ of PATN.}
\begin{table}[t!]
\centering
\setlength{\tabcolsep}{7pt}  
\begin{tabular}{c|ccccc}
\hline
Length $T_{\text{in}}$  & 10    & 20    & 30    & 40    & 50    \\ \hline
ASR(\%) & 34.59 & 37.54 & \textbf{40.11} & 38.81 & 30.88 \\
EER(\%) & 33.34 & 37.84 & \textbf{41.65} & 38.87 & 29.11 \\ \hline
\end{tabular}
\caption{ASR and EER of different input lengths $T_{\text{in}}$ of PATN. We choose $T_{\text{in}}=30$ as our default setting.}
\label{tab:len}
\end{table}


As shown in Table~\ref{tab:len}, PATN's performance is significantly influenced by the input length $T_{\text{in}}$. As the input length increases, PATN captures a longer history of sensor data, improving perturbation accuracy. The optimal performance is observed at an input length of 30, achieving 40.11\% ASR and 41.65\% EER.
However, while longer input lengths allow PATN to leverage more historical data (each unit representing 0.5 seconds of data), excessively long inputs lead to diminishing returns. For input lengths of 40 and 50, performance degrades, suggesting that processing excessive or redundant historical data may dilute the quality of generated perturbations.



\subsubsection{The Impact of History-Aware Top-k Optimization.}

\begin{table}[t!]
\centering
\setlength{\tabcolsep}{9pt} 
\begin{tabular}{l|c|cc}
\hline
Attack  & Alignment & \multicolumn{2}{c}{ Misalignment} \\ 
Method  & PATN               & wHATO\textsuperscript{1}          & w/oHATO\textsuperscript{2}         \\ \hline
ASR(\%) & 40.11              & 39.43              & 30.56                \\
EER(\%) & 41.65              & 40.98              & 33.24                \\ \hline
\end{tabular}
\begin{tablenotes}
\scriptsize
\item \textsuperscript{1} wHATO: with history-aware top-k optimization
\item \textsuperscript{2}  w/oHATO: without history-aware top-k optimization
\end{tablenotes}
\caption{Comparison of ASR and EER with and without History-Aware Top-k Optimization under Misalignment Attacks.}
\label{tab:HATO}
\end{table}

As shown in Table~\ref{tab:HATO}, when subjected to a Misalignment attack with a sliding step length of one second, PATN with HATO achieves 39.43\% ASR and 40.98\% EER. These results demonstrate HATO's effectiveness in addressing Problem 2. HATO mitigates this challenge by dynamically adjusting the perturbation generation process with a history-aware mechanism, optimizing perturbations across multiple sliding windows. This ensures that perturbations align with attack timing, maintaining their effectiveness even when attacks occur at different time steps.


\subsection{Black-box Migration Studies}
\subsubsection{Cross-Input-Length Migration on Privacy Inference Models.}

To assess the transferability and robustness of our adversarial perturbations, we train a single PATN with a fixed output length of $T_{\text{out}} = 10$, and use it to generate adversarial perturbations. These perturbations are then used to attack multiple black-box privacy inference models, each independently trained with a different input length ($T_{\text{priv}} \in {20, 30, 40, 50}$). As shown in Table~\ref{tab:advlen}, the perturbations remain highly effective across all models, with  EER ranging from 38.24\% to 43.02\%. These results demonstrate that our PATN-generated perturbations exhibit strong generalization ability and temporal adaptability, effectively compromising models with varying receptive fields without requiring re-optimization.


\begin{table}[t!]
\centering
\setlength{\tabcolsep}{4.4pt}  
\begin{tabular}{c|ccccc}
\hline
Length  $T_{\text{priv}}$   & 10    & 20    & 30    & 40    & 50    \\ \hline
ASR(\%) & 40.11 & 37.31 & 39.11 & 37.61 & 35.54 \\
Raw data EER(\%) & 8.30 & 7.98 & 6.44 & 5.86 & 5.83\\
Adv data EER(\%) & 41.65 & 41.46 & 43.02 & 41.15 & 38.24
 \\ \hline
\end{tabular}
\caption{ASR and EER of different input lengths  $T_{\text{priv}}$ on privacy inference model.}
\label{tab:advlen}
\end{table}

\begin{table}[t!]
\centering
\setlength{\tabcolsep}{7.8pt} 
\begin{tabular}{c|ccc}
\hline
Black-box model      & MobileNet  & Xception & FCN \\ \hline
ASR(\%)        & 29.43 &   36.57 &  33.48 \\
Raw data EER(\%) &   11.57     &   4.97 &   11.78 \\
Adv data EER(\%) &   36.76      &   37.89 &   38.79\\ \hline
\end{tabular}
\caption{Evaluation of PATN’s effectiveness across multiple black-box architectural models. }
\label{tab:black}
\end{table}
\subsubsection{\textcolor{black}{Diverse Black-Box Privacy Inference Model Architectures.}}


To further evaluate PATN’s transferability, we test its robustness against three unseen black-box privacy inference models: MobileNet, Xception, and a fully convolutional network (FCN). Although trained in a white-box setting with three CNNs (Section~\ref{mumodel}), these models have entirely different architectures. As shown in Table~\ref{tab:black}, PATN maintains consistent performance, achieving ASR of 29.43\%, 36.57\%, and 33.48\%, and EER of 36.76\%, 37.89\%, and 38.79\% on MobileNet, Xception, and FCN, respectively, demonstrating satisfactory privacy protection even against unseen or more complex models.

\section{Conclusion}


In this paper, we propose PATN, a history-driven perturbation generation framework that leverages past sensor data to synthesize future adversarial traces. By conditioning a generative network on historical signals, PATN enables real-time, zero-latency obfuscation while preserving fidelity to the original data. Perturbations can be applied instantly to incoming data, maintaining alignment with natural sensor patterns and mitigating inference-based privacy risks.
For future work, we plan to evaluate PATN in black-box settings across more model architectures and extend its applicability to additional privacy-sensitive attributes, including but not limited to gender and age.

\section{Acknowledgments}
The authors sincerely thank all participants of the peer review. Special appreciation is extended to Xuanqi Gao, Yuhan Zhi, Weipeng Jiang, Yulong Yang, Yuchen Ren, and Chen Ma for their valuable support and insightful suggestions that contributed to the completion of this work.
 This work is supported by the National Key Research and Development Program of China (2023YFB3107401), the National Natural Science Foundation of China (62402377, 62536002, T2341003, 62521002, 62376210, 62161160337, 62132011, U24B20185, U21B2018, 62206217), the Shaanxi Province Key Industry Innovation Program (2023-ZDLGY-38), JSPS KAKENHI JP23K24851, JST PRESTO JPMJPR23P5, JST CREST JPMJCR21M2, JST NEXUS JPMJNX25C4. 
Thanks to the New Cornerstone Science Foundation and the Xplorer Prize.


\bibliography{main}

\setlength{\leftmargini}{20pt}
\makeatletter\def\@listi{\leftmargin\leftmargini \topsep .5em \parsep .5em \itemsep .5em}
\def\@listii{\leftmargin\leftmarginii \labelwidth\leftmarginii \advance\labelwidth-\labelsep \topsep .4em \parsep .4em \itemsep .4em}
\def\@listiii{\leftmargin\leftmarginiii \labelwidth\leftmarginiii \advance\labelwidth-\labelsep \topsep .4em \parsep .4em \itemsep .4em}\makeatother

\setcounter{secnumdepth}{0}
\renewcommand\thesubsection{\arabic{subsection}}
\renewcommand\labelenumi{\thesubsection.\arabic{enumi}}

\newcounter{checksubsection}
\newcounter{checkitem}[checksubsection]

\newcommand{\checksubsection}[1]{%
  \refstepcounter{checksubsection}%
  \paragraph{\arabic{checksubsection}. #1}%
  \setcounter{checkitem}{0}%
}

\newcommand{\checkitem}{%
  \refstepcounter{checkitem}%
  \item[\arabic{checksubsection}.\arabic{checkitem}.]%
}
\newcommand{\question}[2]{\normalcolor\checkitem #1 #2 \color{blue}}
\newcommand{\ifyespoints}[1]{\makebox[0pt][l]{\hspace{-15pt}\normalcolor #1}}

\end{document}